\documentclass[%
 aip,
apl,
amsmath,amssymb,
reprint,%
]{revtex4-1}
\usepackage{hyperref} %adds links
\usepackage{graphicx}% Include figure files
\usepackage{dcolumn}% Align table columns on decimal point
\usepackage{bm}% bold math
\newcommand{\eq}[1]{\begin{equation}#1\end{equation}} % for equation,
\usepackage[mathlines]{lineno}% Enable numbering of text and display math
\usepackage{xcolor} % Add by myself

\begin{document}

	%\preprint{AIP/123-QED}

	\title{Utilization of the Superconducting Transition for Characterizing Low-Quality-Factor Superconducting Resonators}% Force line breaks with \\
	%\thanks{Footnote to title of article.}

	\author{Yu-Cheng Chang}
	\email{randy761005@gate.sinica.edu.tw}
	\affiliation{Department of Physics, National Taiwan University, Taipei 106, Taiwan, Republic of China}%
	\affiliation{Institute of Physics, Academia Sinica, Taipei 115, Taiwan, Republic of China}%
	\affiliation{QTF Centre of Excellence, Department of Applied Physics, Aalto University, P.O.Box 13500, FI-00076 Aalto, Finland}%
	\author{Bayan Karimi}
	\affiliation{QTF Centre of Excellence, Department of Applied Physics, Aalto University, P.O.Box 13500, FI-00076 Aalto, Finland}%
	\author{Jorden Senior}
	\affiliation{QTF Centre of Excellence, Department of Applied Physics, Aalto University, P.O.Box 13500, FI-00076 Aalto, Finland}%
	\author{Alberto Ronzani}
	\affiliation{QTF Centre of Excellence, Department of Applied Physics, Aalto University, P.O.Box 13500, FI-00076 Aalto, Finland}%
	\author{Joonas T. Peltonen}
	\affiliation{QTF Centre of Excellence, Department of Applied Physics, Aalto University, P.O.Box 13500, FI-00076 Aalto, Finland}%
	\author{Hsi-Sheng Goan}
	\affiliation{Department of Physics, National Taiwan University, Taipei 106, Taiwan, Republic of China}%
	\author{Chii-Dong Chen}
	\affiliation{Institute of Physics, Academia Sinica, Taipei 115, Taiwan, Republic of China}%
	\affiliation{QTF Centre of Excellence, Department of Applied Physics, Aalto University, P.O.Box 13500, FI-00076 Aalto, Finland}%
	\author{Jukka P. Pekola}
	\affiliation{QTF Centre of Excellence, Department of Applied Physics, Aalto University, P.O.Box 13500, FI-00076 Aalto, Finland}%

	\date{\today}% It is always \today, today,

	\begin{abstract}

	 Characterizing superconducting microwave resonators with highly dissipative elements is a technical challenge, but a requirement for implementing and understanding the operation of hybrid quantum devices involving dissipative elements, e.g. for thermal engineering and detection. We present experiments on $\lambda/4$ superconducting niobium coplanar waveguide (CPW) resonators, terminating at the antinode by a dissipative copper microstrip via aluminum leads, such that the resonator response is difficult to measure in a typical microwave environment. By measuring the transmission both above and below the superconducting transition of aluminum, we are able to isolate the resonance. We then experimentally verify this method with copper microstrips of increasing thicknesses, from 50 nm to 150 nm, and measure quality factors in the range of $10\sim67$ in a consistent way.

	\end{abstract}

	\maketitle

	\begin{figure}
	\centering
	\includegraphics[width = 1.0 \columnwidth]{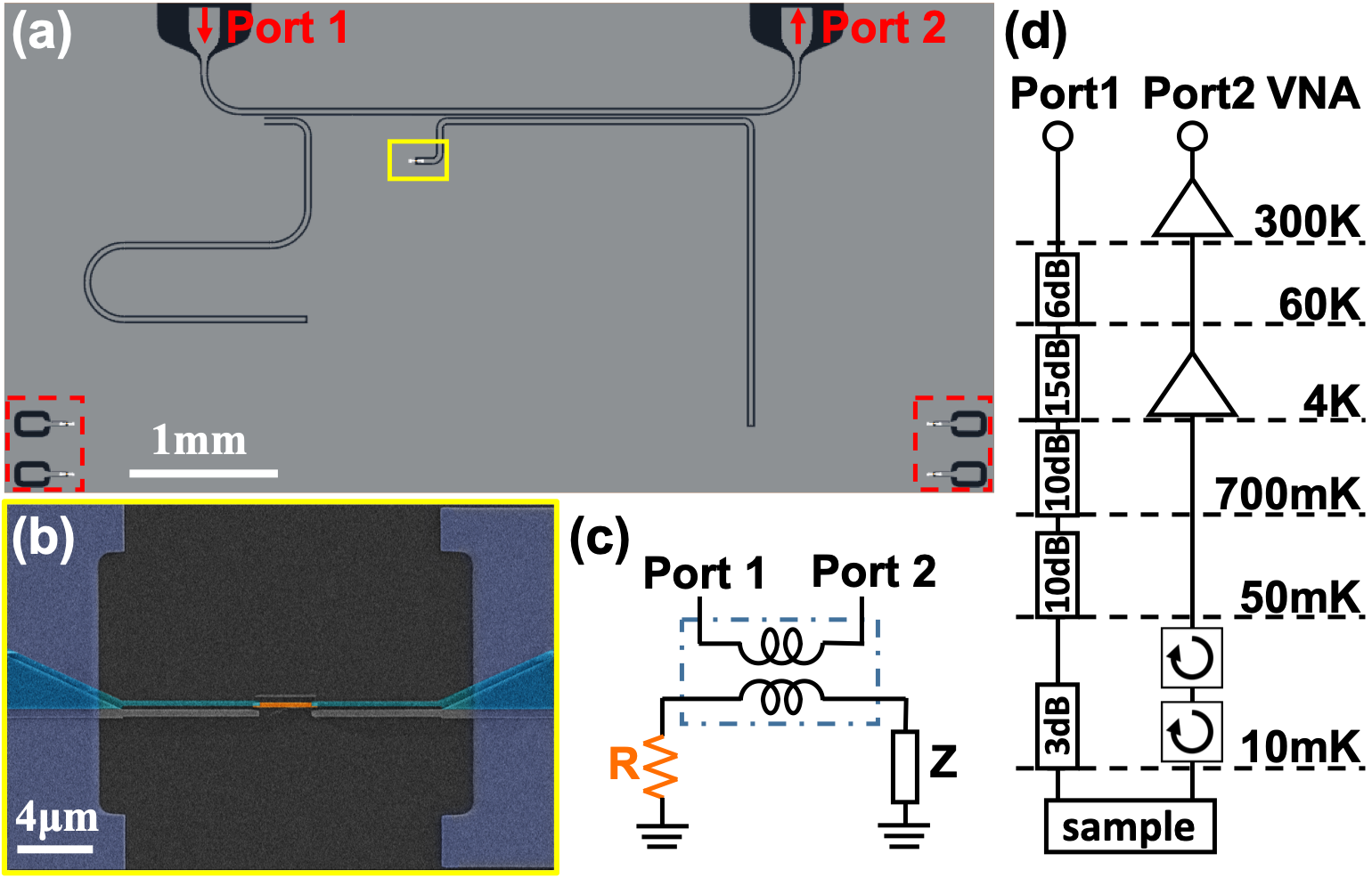}
	\caption{Device and measurement setup (a) The image shows two $\lambda$/4 resonators inductively coupled to a transmission line. The left resonator is a standard $\lambda$/4 superconducting resonator with one end of the center conductor directly shorted to the ground plane on chip, the other end is an open circuit to the ground. This resonator acts as a reference. The right side resonator is a copper-terminated (Al-Cu-Al) $\lambda$/4 resonator. At right and left bottom, four test junctions are made for characterizing the DC electronic properties. (b) Scanning-electron microscope image of yellow outlined region of figure~\ref{device} (a), highlighting Al/Cu/Al junction in contact with the center conductor of resonator at right side and with ground at left side. The image is colored with orange for copper, cyan for aluminum and purple for niobium. (c) Equivalent lumped circuit for copper terminated $\lambda$/4 resonator. The coupled-inductors symbol represents the dominant coupling mechanism between the in/output transmission line and the resonator. The resistance $R$ represents the copper termination. (d) Measurement setup. The microwave signal is introduced from port 1 by a vector network analyzer (VNA) at room temperature and passes through several attenuators distributed at different temperatures to the input of the device at 10 mK. The output microwave is measured at the port 2 of VNA through two isolators and two amplifiers.
	}\label{device}
	\end{figure}

	Superconducting microwave resonators are the cornerstone of much of current state of the art superconducting quantum technologies. Their intrinsic electromagnetic properties routinely enable high quality factors, typically over 10000\cite{Megrant2012}, allowing for extremely sensitive measurements, for example, the multiplexed readout of the dispersive shift of a weakly coupled quantum bit\cite{Majer2007, Chen2012, Jerger2012}, a fundamentally important tool for quantum information, or photon absorption of a microwave kinetic inductance detectors (MKIDs)\cite{Swenson2010, Mchugh2012, Janssen2013, Van2016}, a tool widely used in astronomy. On the other hand, dissipative elements added to superconducting microwave circuits are finding increasing applications in the fields of microwave amplification\cite{Hover2014, Muck2017}, circuit quantum environment engineering\cite{Jones2013,Tuorila2017,Partanen2018}, quantum information\cite{Verstraete2009}, and circuit quantum thermodynamics\cite{Niskanen2007, Karimi2016,Ronzani2018}.

	In superconducting circuits, normal metal elements can be easily integrated into the existing chip architecture, and provide channels for intentional decoherence, such as for quantum bit initialization\cite{Jones2013, Tuorila2017, Partanen2018}, for precise temperature control (electronic heating and cooling)\cite{Nahum1994, Giazotto2006}, ultra sensitive calorimetry\cite{Gasparinetti2015}. Notably, these devices are in use as high-speed thermometers and microwave photon sources\cite{Saira2016, Masuda2018}.

	Experimentally characterizing such dissipative resonators is challenging by conventional single tone spectroscopy, as the dissipation results in a broad and shallow resonance, often unresolvable from the electronic noise of a cryogenic high electron mobility transistor (HEMT) amplifier, and the frequency dependent variations of transmission in the experimental setup. In this report, we present a method for this task utilizing an intermediate superconductor with a lower critical temperature, enabling us to isolate the resonance by performing characterization at differing bath temperatures, with measured quality factors as low as 10.

%Sample

	The method presented is demonstrated here using a structure consisting of two $\lambda/4$ niobium CPW resonators, both inductively coupled to a common CPW transmission line, shown in Fig.~\ref{device} (a), used for multiplexed readout. On the left is a fully superconducting $\lambda/4$ CPW resonator to act as a reference. On the right resonator, the voltage node is terminated by an aluminum-copper-aluminum constriction, forming a Nb/Al/Cu/Al/Nb heterostructure, with the copper acting as the dissipative normal metal shown in the micrograph in Fig.~\ref{device} (b).

	Samples are fabricated on a 330 $\mu$m thick c-plane sapphire substrate using a process described in Ref.~\onlinecite{Ronzani2018}. The sapphire surface is initially cleaned with an argon ion plasma milling before a 200 nm thick niobium film is deposited by DC magnetron sputtering. The coplanar waveguide resonator patterns are written by electron-beam lithography (EBL), and transferred to the niobium film using an SF$_6$ + O$_2$ reactive ion etching process. As the niobium layer is thick, subsequently comparatively thin evaporated aluminum and copper films are fragile and can become discontinuous at the intersection. Therefore, during the EBL exposure, the dose at the interface is incrementally changed, forming a ramp when etched to remove the discontinuity and increase the surface area.  The Al/Cu/Al terminations are written by electron-beam lithography onto a bilayer resist and grown using double-angle deposition in an electron-beam evaporator, with galvanic contact between lithographic layers facilitated by an $in$-$situ$ argon ion plasma milling process removing native oxides on the niobium. At the beginning of deposition, 10 nm of aluminum is evaporated to improve adhesion to the sapphire surface, followed by copper (of variable thickness; 50 nm, 100 nm, and 150 nm for different devices). Finally, two 110 nm thick aluminum contacts to the niobium CPW are deposited. After processing, the substrates are diced with a diamond-embedded resin blade, wire-bonded to the sample stage assembly described in Ref.~\onlinecite{George2017}. Finally, the sample is loaded into a cryogen-free dilution refrigerator with a base temperature of 10 mK.
%Setup

	All spectroscopic measurements are performed using a vector network analyzer (VNA) at room temperature with the signal reaching the sample via an attenuated microwave line. The attenuation is distributed at various temperature stages of the dilution refrigerator, as shown in Fig.~\ref{device} (d). This set-up reduces Johnson-Nyquist noise from the attenuators at higher temperature stages\cite{Urick2015}. The signal leaving the sample is then passed through two circulators at base temperature, to a low noise HEMT amplifier mounted at 4 K, providing 80 dB isolation from the amplifier input. Outside of the dilution refrigerator, the signal is passed through an additional 30 dB amplifier, before it is received by the VNA, capturing both $I$ and $Q$ quadratures, from which we reconstitute the transmission $S_{21}$. The transmission through the system reads

	\eq{
		S_{21}^{notch}=ae^{i\alpha}e^{-2\pi if\tau}[1-\frac{ (Q_l/|Q_c|)e^{i\phi}}{1+2iQ_l(f/f_r-1)}].
		\label{sparameter}
	}

	Here, $a$ is the overall amplitude, $\alpha$ is the phase shift contributed by various components in the circuit, $\tau$ is the electronic delay caused by the length of the cable and the finite speed of light, and $\phi$ quantifies the impedance mismatch\cite{Khalil2012, Deng2013}. Parameters $f$ and $f_r$ denote the probe frequency and the resonance frequency of the resonator, and $Q_l$ and $Q_c$ are the loaded quality factor and the coupling quality factor, respectively\cite{Probst2015}. The quality factor is the ratio of energy stored in a resonator to average energy lost per cycle. The photon loss rate, inversely proportional to the quality factor, is a linear combination of the internal losses of the resonator, and corresponds to the losses arising in the coupling from the resonator to the transmission line. The inverse of the loaded quality factor reads $1/Q_l=1/Q_i+1/Q_c$. The $Q_l/Q_c$ ratio determines the depth of the notched transmission and is maximized at $Q_l\approx Q_c$ at the resonance frequency. In order to increase the measured signal for a lossy resonator, we require $Q_c\ll Q_i$, meaning that there are more photons leaving the resonator than dissipated ones.

	In the samples measured, a quality factor of 20 was desired corresponding to favorable operation of the envisioned device, the quantum heat engine\cite{Karimi2016, Ronzani2018}. The quality factor of our samples can be estimated by a simple model which terminates the center conductor of the CPW transmission line and the ground through a lumped resistance $R$ at the voltage node, and capacitively to the ground at the current node. The impedance of our $\lambda/4$ transmission line, shown in Fig.~\ref{device} (c), is given near this resonance by $Z=-i\pi Z_{\infty}(f/f_r-f_r/f)/4$, where $Z_{\infty}$= 50 $\Omega$ is the characteristic impedance of an infinite transmission line. Power injected to the termination at the half-power points $f=f_r \pm\Delta f/2$ can be written as $P \simeq V^2/ ( R (1+ ( \pi Z_{\infty} / 4R)^2(\Delta f/f_r)^2 ) )=V^2/2R$, where $\Delta f$ is the width of the resonance peak. By definition, the internal quality factor $Q_i$ of the resistively terminated resonator can be written
as
	\eq{
		Q_i=\frac{\pi Z_{\infty}}{4R}.
		\label{quality factor}
	}

\begin{figure}
	\centering
	\includegraphics[width = 1.0\columnwidth]{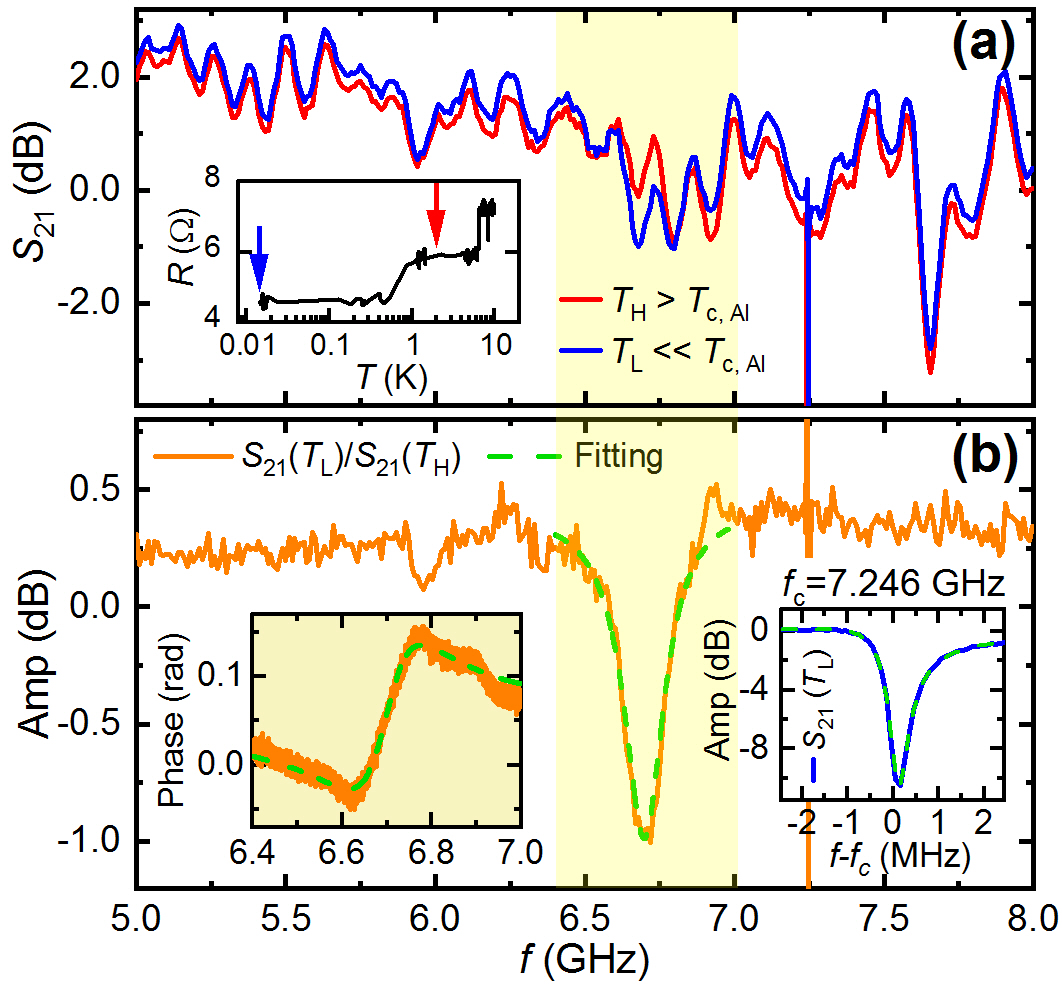}
	\caption{(a) Transmittance spectra of the device with 100 nm thick copper termination. The red and blue curves are measured at $T_{H}$ = 2 K and $T_{L}$ = 10 mK respectively, as indicated with arrows in the inset. Here, the $T_{L}$ and $T_{H}$ satisfy $T_{L}\ll T_{c,Al}<T_{H}<T_{c,Nb}$. The inset shows resistance-temperature characteristics of a nominally identical co-process dissipative element, demonstrating three plateaus corresponding to the superconducting transitions of aluminum, and niobium, respectively. The yellow region demonstrates the frequency range of the resistively terminated superconducting resonator. We present in the right inset of (b) the quality factor of $2.8\times10^4$ at resonant frequency $f_0 =7.246$ GHz of the reference resonator. (b) The amplitude of $S_{21}(T_L)/S_{21}(T_H)$ and the phase in the left inset. Within the yellow highlighted frequency region, we see the resonance of the dissipative resonator. Here, we extracted $f_r$ and $Q_i$ as 6.71 GHz and 45 corresponding to the green dashed fitting line.
	}\label{spectroscopy}
	\end{figure}

	To measure the low $Q_i$, coupling should be increased ($Q_c\ll Q_i$)\cite{Goppl2008}, in order to maximize the measurable signal. For us, however, even though the coupler covers half of the cavity length to achieve the strong coupling regime, the $Q_l$ is still dominated by $Q_i$. Due to the low overall quality factor, the depth of the notch in $S_{21}$ is of order 1 dB, which typically cannot be resolved within the background of the microwave setup. To isolate the resonator, the method presented here is based on measuring the spectrum above and below the critical temperature of aluminum $T_{c, Al}$. The lower quality factor $Q_i$ of the resonator above $T_{c, Al}$ suppresses the resonance. Under these “high temperature” conditions, the transmittance can be regarded as a reference measurement for the background. The transmission through the CPW line is shown in Fig.~\ref{spectroscopy} (b), with the red and blue traces measured at different temperatures $T_H$ and $T_L$, above and below the critical temperature of the aluminum $T_{c, Al}$, respectively, yet staying sufficiently below the critical temperature $T_{c, Nb}$ of the niobium transmission line. Here, $T_{c, Al}$ and $T_{c, Nb}$ can be determined by the resistance-temperature characteristic of a co-process sample shown in the inset of Fig.~\ref{spectroscopy} (a). A narrow band of $S_{21}(T_L)$ is shown in the right inset of Fig.~\ref{spectroscopy} (b) emphasising the narrow bandwidth of the standard reference resonator. Here, $f_0$ is extracted as 7.246 GHz with a corresponding 5 MHz shift with temperature, as the kinetic inductance increases with decreasing temperature\cite{Meservey1969, Inomata2009}. The dissipative resonator is not clearly visible in either trace. We observe, however, that as expected, due to lower operating temperature and thus more ideal superconducting characteristics of the CPW transmission line, the blue trace is consistently higher than the red trace, except in the region highlighted in yellow, centered around the design frequency of the dissipative resonator. By taking the ratio of the two $S_{21}$ measured at $T_L$ and $T_H$, one can isolate the resonance of the dissipative resonator. This ratio is shown in Fig.~\ref{spectroscopy} (b). Fitting a standard notched resonator model\cite{Probst2015} to the trace, the parameters $f_r$, $Q_c$, and $Q_i$ can be extracted, with data for different samples (with variable thickness of copper) presented in Table~\ref{rfresult}.

%Measurements

\begin{table}[h]
\caption{Parameters of resistively terminated superconducting resonators extracted based on the measured $S_{21}$.  Here,  \textdagger\  refers to samples with increased contact resistance between the niobium and aluminum layers.}
\centering
\begin{tabular}{l|c|ccc|c|cc}

\hline\hline
Thickness		& 50 nm	&  &100 nm&	&150nm&\multicolumn{2}{c}{\centering \textdagger100 nm}\\

\hline           % inserts single
sample				& A-1 	& B-1 	& B-2 	& B-3	& C-1	& D-1	& D-2\\
$f_{r}$ (GHz) 			& 6.54 	& 6.74	&6.69	&  6.70	& 6.53	& 6.67	& 6.67\\
$Q_c$				& 330	& 340	& 200 	& 250	& 320	& 150	& 460\\
$Q_{i}$				& 10.3	& 45.3	& 53.2	& 44.7	& 66.3	& 20.4	& 26.5\\
$Q_{i, err}$			& 0.5		& 0.5		& 0.9		& 0.3		& 1.3		& 0.2		& 0.8\\

\hline\hline

\end{tabular}
\label{rfresult}
\end{table}

\begin{figure}
	\centering
	\includegraphics[width = 0.9\columnwidth]{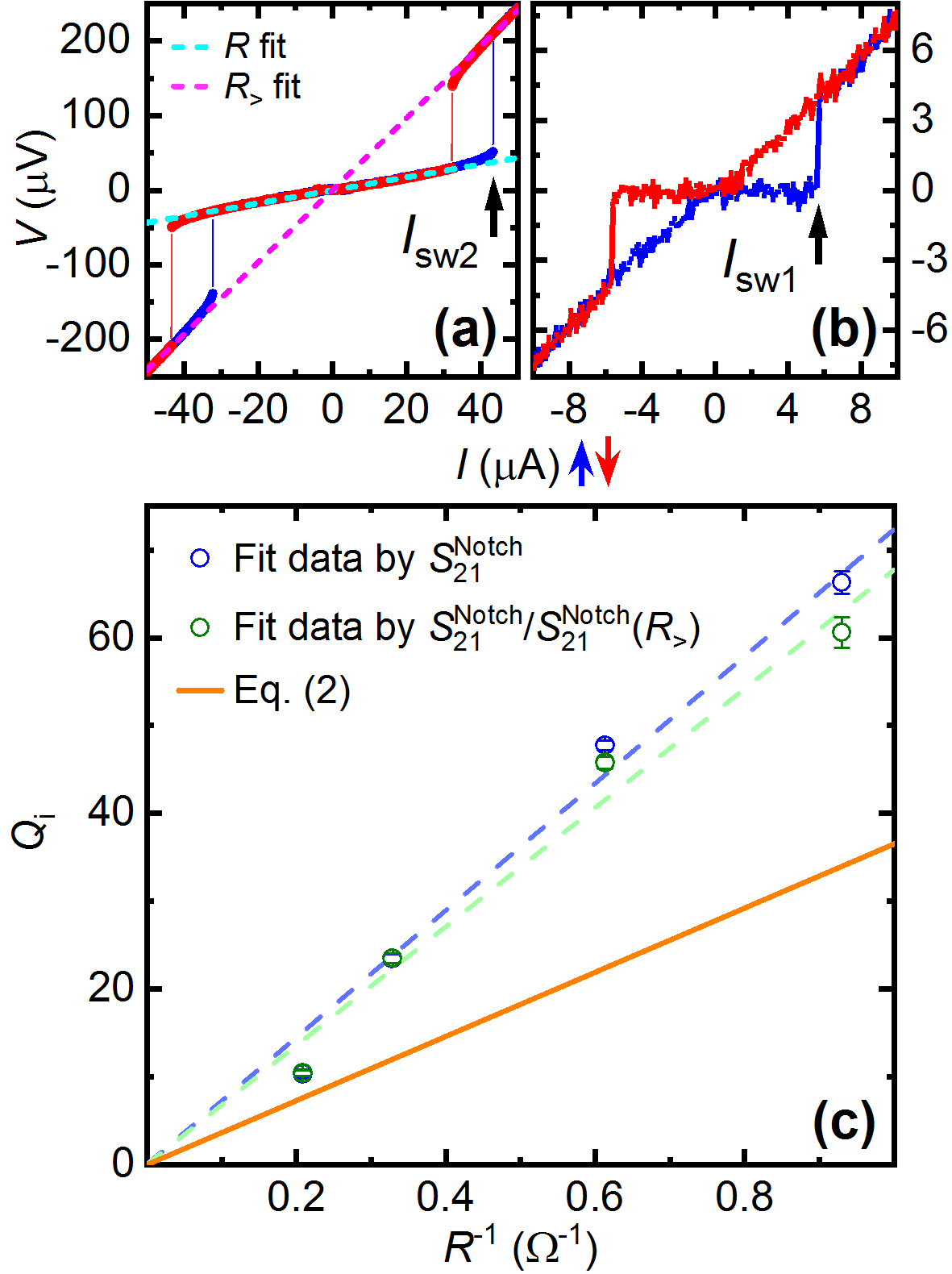}
	\caption{(a)(b) $VI$ measurement for termination element of Nb/Al/Cu/Al/Nb of the 150 nm thick copper. $I_{sw1}$ is the critical current of the SNS junction and $I_{sw2}$ denotes the critical current of aluminum. Applying current below $I_{sw2}$ and over $I_{sw1}$ leads to the slope marked with light blue dashed line originating from the resistance $R$ of copper wire and the Cu/Al contact resistance. At a bias current exceeding the critical current of the aluminum wire, the copper, aluminum, Cu/Al contact resistance, and Nb/Al contact resistance, all contribute to the total resistance $R_{>}$. (c) $Q_i$ extracted by fitting two models, notch $S_{21}$ and divided notch $S_{21}/S_{21}(R_>)$ versus the reciprocal measured termination resistance $R$ in blue dots and green hollow circles separately. Orange solid line denotes theoretical $Q_i$ as a function of reciprocal termination resistance.
	}\label{compare}
	\end{figure}

	In order to investigate how the $Q_i$ of the resistively terminated superconducting resonator depends on its resistance $R$, the co-process samples with identical termination element (see the red dashed frame in Fig.~\ref{device} (a)) are measured in current biased four probe configuration at 50 mK bath temperature. A measured $IV$ curve is shown for the 150 nm thick copper termination in Fig.~\ref{compare} (a). By ramping the bias current up through the termination, we observe first the superconducting energy gap and then two resistive branches at higher bias. The superconducting state exists by virtue of the proximity effect in the SNS Josephson junction induced by the Al/Cu/Al structure\cite{Wei2011, Angers2008}, and the termination switches to the first dissipative branch at current $I_{sw1}$ = 5.6 $\mu$A with normal state resistance $R$ = 0.87 $\Omega$. The normal resistance $R$ due to the copper wire and the imperfect contact between the aluminum and the copper layers determines the quality factor of the resonator in the microwave measurement at $T_L$. Continually increasing the current, the resistance increases at $I_{sw2}$ = 43.3 $\mu$A with resistance $R_{>}$ = 4.83 $\Omega$ when the current exceeds the critical current of aluminum. $R_{>}$ consists of $R$, the normal state resistance of aluminum wire and the Al/Nb contact. All the measured parameters $I_{sw1}$, $I_{sw2}$, $R$ and $R_{>}$ extracted from $IV$ measurement are given in Table~\ref{dcresult}.

\begin{table}[h]
\caption{DC properties of Nb/Al/Cu/Al/Nb junctions.}
\centering
\begin{tabular}{l|rr|rr|rr|rr}
\hline\hline
Thickness  &\multicolumn{2}{c|}{\centering 50 nm} &\multicolumn{2}{c|}{\centering 100 nm} &\multicolumn{2}{c|}{\centering 150 nm} &\multicolumn{2}{c}{\centering \textdagger100 nm}\\

\hline           % inserts single
sample					& A-2		& A-3		& B-4 		& B-5		& C-2		& C-3		& D-3		& D-4\\
$I_{sw1}$ ($\mu$A) 	& 0.2 	& 0.3		& 2.1		& 1.8		& 5.6		& 2.6		& 0.74	& 1.0\\
$I_{sw2}$ ($\mu$A) 	& 19.6	& 22.8	& 39.0	& 40.0	& 43.3	& 42.0	& 21.0	& 21.1\\
$R$ ($\Omega$)	& 5.3		&4.3		& 1.7		& 1.6		& 0.9		& 1.3		&3.3		& 2.8\\
$R_{>}$ ($\Omega$)		& 21.5	&16.0	& 8.2		& 7.9		&4.8	&5.2	&19.4	& 19.5\\

\hline\hline
\end{tabular}
\label{dcresult}
\end{table}

	The presence of supercurrent in SNS junctions is due to the well-known proximity effect\cite{Clarke1969, Shepherd1972, DeGennes1964, Dubos2001}. Based on the ratio of the Thouless energy $E_T=\hbar D/L^{2}$ to the superconducting gap of aluminum $\Delta$, the junction is in the long junction regime, when $E_T\Delta \ll 1$ and the zero temperature $eRI_{c}$ is found to be proportional to $E_T$ in this limit. Here $L$ is the length of the junction, $D=v_{F}l_{e}/3$ is the diffusion constant of the N metal, $v_{F}$ is the Fermi velocity, $I_{c}$ is the critical current of the junction, and $l_e$ is the elastic mean free path of electrons. In Table~\ref{dcresult}, $eRI_{sw1}$ is smaller than 5 $\mu$eV indicating long junction limit.

%Fermi velocity: http://hyperphysics.phy-astr.gsu.edu/hbase/Tables/fermi.html

	The N lead between S superconductors of the resonator is an ideal element to localise heat. The small volume of normal metal enhances the temperature of the element at fixed transferred power, whereas the S forms a Cooper-pair while injecting an electron from N and reflects a hole known as Andreev reflection that drops the temperature at the interface and localises the heat efficiently in the N\cite{Andreev1965}. Depending on the application and operation regime, the quality factor $Q_i$ of the resistively terminated resonator can be designed merely by the resistance of the N element in the long junction regime. In Ref.~\onlinecite{Ronzani2018}, $f_{r}/Q_i$ is identified as the coupling between the copper heat bath and the resonator.

	In Fig.~\ref{compare} (c), we plot the measured $Q_i$ from Table~\ref{rfresult} as blue dots. The theoretically expected $Q_i$ is shown as orange line. The expected slope by Eq.~(\ref{quality factor}) is shown with $Z_{\infty}$ set to 50 $\Omega$. For a more general analysis of the internal quality factor, a model that takes into account non-vanishing quality factor above the aluminum superconducting transition temperature can be used to simulate the measurement technique by inserting $R_{>}$ into Eq.~(\ref{quality factor}) to obtain $Q_i(T_{H})$. $Q_i(T_{L})$ and $Q_i(T_{H})$ are independent dissipation channels with $1/Q_{l}=1/Q_i+1/Q_{c}$, which are then inserted into Eq.~(\ref{sparameter}) to obtain the general two temperature model, $S_{21}(T_{L})/S_{21}(T_{H})$. Here $Q_i(T_{L})$ is extracted from the model and plotted as green hollow circles. The slopes of extracted $Q_i$ vs $1/R$ from the two models differ by less than 12\%, which places the values within the uncertainty of the fitting algorithm. According to Table~\ref{dcresult}, $R_{>}$ decreases while the thickness of the copper increases. The lower $R_{>}$ in the thicker samples yields a broader resonance when aluminum undergoes the transition to normal state, which explains the difference of extracted $Q_i$ for the thicker Cu films. However, the measured $Q_i$ is approximately double the expected value. The plausible origin of the discrepancy with respect to Eq.~(\ref{quality factor}) is that admittance of the SNS junction is composed of parallel dissipative (real) and reactive (imaginary) components, $Y_R$ and $Y_I$, respectively~\cite{Virtanen2011}. In this case the actual quality factor $Q_i$ is enhanced with respect to Eq.~(\ref{quality factor}) by a factor $(Y_R^2+Y_I^2)/Y_R^2$, and while the contribution of the phase is challenging to measure, based on the $I_{sw1}$ and $R$ values presented in Table ~\ref{dcresult}, we estimate that this can lead to an enhancement of approximately factor 2, which would explain the discrepancy.

%What's different between supercurrent and switch current?
	In conclusion, we have presented a technique for isolating the resonance of low-quality dissipative superconducting resonators based on the transition temperature of an intermediate superconductor with a lower energy gap as compared to that of the superconducting resonator. We have verified this method by characterizing a series of resistively terminated $\lambda/4$ superconducting CPW resonators. From microwave measurements, we have extracted extremely low quality factors of $10-67$ for resonators with different resistances of copper termination element. While similar characterization exploiting the superconducting transition could be achieved by supplying sufficient power to the resonator in excess of the critical current of aluminum, or bypassed completely by using a microwave switch to measure a background reference, we believe the technique presented here is more versatile, applicable also in highly attenuated microwave input lines, such as those being used in circuit quantum electrodynamics and thermodynamics experiments, and more accurately measuring the microwave background closer to the device being characterized.

	We acknowledge Micronova Nanofabrication Centre of OtaNano infrastructure for providing the processing facilities, and the European MicroKelvin Platform for the use of the dilution refrigerator. Y.-C.C. is supported by the visiting project of the Centre for Quantum Engineering, Finland and Institute of Physics, Academia Sinica, Taiwan (R.O.C.) during this experiment. B. K. acknowledges the grant by Marie Sklodowska-Curie actions (grant agreements 742559 and 766025).

% Insert bibliography

\bibliography{bibtex}

\begin{thebibliography}{38}%
\makeatletter
\providecommand \@ifxundefined [1]{%
 \@ifx{#1\undefined}
}%
\providecommand \@ifnum [1]{%
 \ifnum #1\expandafter \@firstoftwo
 \else \expandafter \@secondoftwo
 \fi
}%
\providecommand \@ifx [1]{%
 \ifx #1\expandafter \@firstoftwo
 \else \expandafter \@secondoftwo
 \fi
}%
\providecommand \natexlab [1]{#1}%
\providecommand \enquote  [1]{``#1''}%
\providecommand \bibnamefont  [1]{#1}%
\providecommand \bibfnamefont [1]{#1}%
\providecommand \citenamefont [1]{#1}%
\providecommand \href@noop [0]{\@secondoftwo}%
\providecommand \href [0]{\begingroup \@sanitize@url \@href}%
\providecommand \@href[1]{\@@startlink{#1}\@@href}%
\providecommand \@@href[1]{\endgroup#1\@@endlink}%
\providecommand \@sanitize@url [0]{\catcode `\\12\catcode `\$12\catcode
  `\&12\catcode `\#12\catcode `\^12\catcode `\_12\catcode `\%12\relax}%
\providecommand \@@startlink[1]{}%
\providecommand \@@endlink[0]{}%
\providecommand \url  [0]{\begingroup\@sanitize@url \@url }%
\providecommand \@url [1]{\endgroup\@href {#1}{\urlprefix }}%
\providecommand \urlprefix  [0]{URL }%
\providecommand \Eprint [0]{\href }%
\providecommand \doibase [0]{http://dx.doi.org/}%
\providecommand \selectlanguage [0]{\@gobble}%
\providecommand \bibinfo  [0]{\@secondoftwo}%
\providecommand \bibfield  [0]{\@secondoftwo}%
\providecommand \translation [1]{[#1]}%
\providecommand \BibitemOpen [0]{}%
\providecommand \bibitemStop [0]{}%
\providecommand \bibitemNoStop [0]{.\EOS\space}%
\providecommand \EOS [0]{\spacefactor3000\relax}%
\providecommand \BibitemShut  [1]{\csname bibitem#1\endcsname}%
\let\auto@bib@innerbib\@empty
%</preamble>
\bibitem [{\citenamefont {Megrant}\ \emph {et~al.}(2012)\citenamefont
  {Megrant}, \citenamefont {Neill}, \citenamefont {Barends}, \citenamefont
  {Chiaro}, \citenamefont {Chen}, \citenamefont {Feigl}, \citenamefont {Kelly},
  \citenamefont {Lucero}, \citenamefont {Mariantoni}, \citenamefont {O'Malley}
  \emph {et~al.}}]{Megrant2012}%
  \BibitemOpen
  \bibfield  {author} {\bibinfo {author} {\bibfnamefont {A.}~\bibnamefont
  {Megrant}}, \bibinfo {author} {\bibfnamefont {C.}~\bibnamefont {Neill}},
  \bibinfo {author} {\bibfnamefont {R.}~\bibnamefont {Barends}}, \bibinfo
  {author} {\bibfnamefont {B.}~\bibnamefont {Chiaro}}, \bibinfo {author}
  {\bibfnamefont {Y.}~\bibnamefont {Chen}}, \bibinfo {author} {\bibfnamefont
  {L.}~\bibnamefont {Feigl}}, \bibinfo {author} {\bibfnamefont
  {J.}~\bibnamefont {Kelly}}, \bibinfo {author} {\bibfnamefont
  {E.}~\bibnamefont {Lucero}}, \bibinfo {author} {\bibfnamefont
  {M.}~\bibnamefont {Mariantoni}}, \bibinfo {author} {\bibfnamefont {P.~J.}\
  \bibnamefont {O'Malley}},  \emph {et~al.},\ }\href@noop {} {\bibfield
  {journal} {\bibinfo  {journal} {Appl. Phys. Lett.}\ }\textbf {\bibinfo
  {volume} {100}},\ \bibinfo {pages} {113510} (\bibinfo {year}
  {2012})}\BibitemShut {NoStop}%
\bibitem [{\citenamefont {Majer}\ \emph {et~al.}(2007)\citenamefont {Majer},
  \citenamefont {Chow}, \citenamefont {Gambetta}, \citenamefont {Koch},
  \citenamefont {Johnson}, \citenamefont {Schreier}, \citenamefont {Frunzio},
  \citenamefont {Schuster}, \citenamefont {Houck}, \citenamefont {Wallraff}
  \emph {et~al.}}]{Majer2007}%
  \BibitemOpen
  \bibfield  {author} {\bibinfo {author} {\bibfnamefont {J.}~\bibnamefont
  {Majer}}, \bibinfo {author} {\bibfnamefont {J.}~\bibnamefont {Chow}},
  \bibinfo {author} {\bibfnamefont {J.}~\bibnamefont {Gambetta}}, \bibinfo
  {author} {\bibfnamefont {J.}~\bibnamefont {Koch}}, \bibinfo {author}
  {\bibfnamefont {B.}~\bibnamefont {Johnson}}, \bibinfo {author} {\bibfnamefont
  {J.}~\bibnamefont {Schreier}}, \bibinfo {author} {\bibfnamefont
  {L.}~\bibnamefont {Frunzio}}, \bibinfo {author} {\bibfnamefont
  {D.}~\bibnamefont {Schuster}}, \bibinfo {author} {\bibfnamefont
  {A.}~\bibnamefont {Houck}}, \bibinfo {author} {\bibfnamefont
  {A.}~\bibnamefont {Wallraff}},  \emph {et~al.},\ }\href@noop {} {\bibfield
  {journal} {\bibinfo  {journal} {Nature}\ }\textbf {\bibinfo {volume} {449}},\
  \bibinfo {pages} {443} (\bibinfo {year} {2007})}\BibitemShut {NoStop}%
\bibitem [{\citenamefont {Chen}\ \emph {et~al.}(2012)\citenamefont {Chen},
  \citenamefont {Sank}, \citenamefont {O'Malley}, \citenamefont {White},
  \citenamefont {Barends}, \citenamefont {Chiaro}, \citenamefont {Kelly},
  \citenamefont {Lucero}, \citenamefont {Mariantoni}, \citenamefont {Megrant}
  \emph {et~al.}}]{Chen2012}%
  \BibitemOpen
  \bibfield  {author} {\bibinfo {author} {\bibfnamefont {Y.}~\bibnamefont
  {Chen}}, \bibinfo {author} {\bibfnamefont {D.}~\bibnamefont {Sank}}, \bibinfo
  {author} {\bibfnamefont {P.}~\bibnamefont {O'Malley}}, \bibinfo {author}
  {\bibfnamefont {T.}~\bibnamefont {White}}, \bibinfo {author} {\bibfnamefont
  {R.}~\bibnamefont {Barends}}, \bibinfo {author} {\bibfnamefont
  {B.}~\bibnamefont {Chiaro}}, \bibinfo {author} {\bibfnamefont
  {J.}~\bibnamefont {Kelly}}, \bibinfo {author} {\bibfnamefont
  {E.}~\bibnamefont {Lucero}}, \bibinfo {author} {\bibfnamefont
  {M.}~\bibnamefont {Mariantoni}}, \bibinfo {author} {\bibfnamefont
  {A.}~\bibnamefont {Megrant}},  \emph {et~al.},\ }\href@noop {} {\bibfield
  {journal} {\bibinfo  {journal} {Appl. Phys. Lett.}\ }\textbf {\bibinfo
  {volume} {101}},\ \bibinfo {pages} {182601} (\bibinfo {year}
  {2012})}\BibitemShut {NoStop}%
\bibitem [{\citenamefont {Jerger}\ \emph {et~al.}(2012)\citenamefont {Jerger},
  \citenamefont {Poletto}, \citenamefont {Macha}, \citenamefont {H{\"u}bner},
  \citenamefont {Il’ichev},\ and\ \citenamefont {Ustinov}}]{Jerger2012}%
  \BibitemOpen
  \bibfield  {author} {\bibinfo {author} {\bibfnamefont {M.}~\bibnamefont
  {Jerger}}, \bibinfo {author} {\bibfnamefont {S.}~\bibnamefont {Poletto}},
  \bibinfo {author} {\bibfnamefont {P.}~\bibnamefont {Macha}}, \bibinfo
  {author} {\bibfnamefont {U.}~\bibnamefont {H{\"u}bner}}, \bibinfo {author}
  {\bibfnamefont {E.}~\bibnamefont {Il’ichev}}, \ and\ \bibinfo {author}
  {\bibfnamefont {A.~V.}\ \bibnamefont {Ustinov}},\ }\href@noop {} {\bibfield
  {journal} {\bibinfo  {journal} {Appl. Phys. Lett.}\ }\textbf {\bibinfo
  {volume} {101}},\ \bibinfo {pages} {042604} (\bibinfo {year}
  {2012})}\BibitemShut {NoStop}%
\bibitem [{\citenamefont {Swenson}\ \emph {et~al.}(2010)\citenamefont
  {Swenson}, \citenamefont {Cruciani}, \citenamefont {Benoit}, \citenamefont
  {Roesch}, \citenamefont {Yung}, \citenamefont {Bideaud},\ and\ \citenamefont
  {Monfardini}}]{Swenson2010}%
  \BibitemOpen
  \bibfield  {author} {\bibinfo {author} {\bibfnamefont {L.}~\bibnamefont
  {Swenson}}, \bibinfo {author} {\bibfnamefont {A.}~\bibnamefont {Cruciani}},
  \bibinfo {author} {\bibfnamefont {A.}~\bibnamefont {Benoit}}, \bibinfo
  {author} {\bibfnamefont {M.}~\bibnamefont {Roesch}}, \bibinfo {author}
  {\bibfnamefont {C.}~\bibnamefont {Yung}}, \bibinfo {author} {\bibfnamefont
  {A.}~\bibnamefont {Bideaud}}, \ and\ \bibinfo {author} {\bibfnamefont
  {A.}~\bibnamefont {Monfardini}},\ }\href@noop {} {\bibfield  {journal}
  {\bibinfo  {journal} {Appl. Phys. Lett.}\ }\textbf {\bibinfo {volume} {96}},\
  \bibinfo {pages} {263511} (\bibinfo {year} {2010})}\BibitemShut {NoStop}%
\bibitem [{\citenamefont {McHugh}\ \emph {et~al.}(2012)\citenamefont {McHugh},
  \citenamefont {Mazin}, \citenamefont {Serfass}, \citenamefont {Meeker},
  \citenamefont {O'Brien}, \citenamefont {Duan}, \citenamefont {Raffanti},\
  and\ \citenamefont {Werthimer}}]{Mchugh2012}%
  \BibitemOpen
  \bibfield  {author} {\bibinfo {author} {\bibfnamefont {S.}~\bibnamefont
  {McHugh}}, \bibinfo {author} {\bibfnamefont {B.~A.}\ \bibnamefont {Mazin}},
  \bibinfo {author} {\bibfnamefont {B.}~\bibnamefont {Serfass}}, \bibinfo
  {author} {\bibfnamefont {S.}~\bibnamefont {Meeker}}, \bibinfo {author}
  {\bibfnamefont {K.}~\bibnamefont {O'Brien}}, \bibinfo {author} {\bibfnamefont
  {R.}~\bibnamefont {Duan}}, \bibinfo {author} {\bibfnamefont {R.}~\bibnamefont
  {Raffanti}}, \ and\ \bibinfo {author} {\bibfnamefont {D.}~\bibnamefont
  {Werthimer}},\ }\href@noop {} {\bibfield  {journal} {\bibinfo  {journal}
  {Rev. Sci. Instrum.}\ }\textbf {\bibinfo {volume} {83}},\ \bibinfo {pages}
  {044702} (\bibinfo {year} {2012})}\BibitemShut {NoStop}%
\bibitem [{\citenamefont {Janssen}\ \emph {et~al.}(2013)\citenamefont
  {Janssen}, \citenamefont {Baselmans}, \citenamefont {Endo}, \citenamefont
  {Ferrari}, \citenamefont {Yates}, \citenamefont {Baryshev},\ and\
  \citenamefont {Klapwijk}}]{Janssen2013}%
  \BibitemOpen
  \bibfield  {author} {\bibinfo {author} {\bibfnamefont {R.}~\bibnamefont
  {Janssen}}, \bibinfo {author} {\bibfnamefont {J.}~\bibnamefont {Baselmans}},
  \bibinfo {author} {\bibfnamefont {A.}~\bibnamefont {Endo}}, \bibinfo {author}
  {\bibfnamefont {L.}~\bibnamefont {Ferrari}}, \bibinfo {author} {\bibfnamefont
  {S.}~\bibnamefont {Yates}}, \bibinfo {author} {\bibfnamefont
  {A.}~\bibnamefont {Baryshev}}, \ and\ \bibinfo {author} {\bibfnamefont
  {T.}~\bibnamefont {Klapwijk}},\ }\href@noop {} {\bibfield  {journal}
  {\bibinfo  {journal} {Appl. Phys. Lett.}\ }\textbf {\bibinfo {volume}
  {103}},\ \bibinfo {pages} {203503} (\bibinfo {year} {2013})}\BibitemShut
  {NoStop}%
\bibitem [{\citenamefont {van Rantwijk}\ \emph {et~al.}(2016)\citenamefont {van
  Rantwijk}, \citenamefont {Grim}, \citenamefont {van Loon}, \citenamefont
  {Yates}, \citenamefont {Baryshev},\ and\ \citenamefont
  {Baselmans}}]{Van2016}%
  \BibitemOpen
  \bibfield  {author} {\bibinfo {author} {\bibfnamefont {J.}~\bibnamefont {van
  Rantwijk}}, \bibinfo {author} {\bibfnamefont {M.}~\bibnamefont {Grim}},
  \bibinfo {author} {\bibfnamefont {D.}~\bibnamefont {van Loon}}, \bibinfo
  {author} {\bibfnamefont {S.}~\bibnamefont {Yates}}, \bibinfo {author}
  {\bibfnamefont {A.}~\bibnamefont {Baryshev}}, \ and\ \bibinfo {author}
  {\bibfnamefont {J.}~\bibnamefont {Baselmans}},\ }\href@noop {} {\bibfield
  {journal} {\bibinfo  {journal} {IEEE Trans. Microw. Theory Tech.}\ }\textbf
  {\bibinfo {volume} {64}},\ \bibinfo {pages} {1876} (\bibinfo {year}
  {2016})}\BibitemShut {NoStop}%
\bibitem [{\citenamefont {Hover}\ \emph {et~al.}(2014)\citenamefont {Hover},
  \citenamefont {Zhu}, \citenamefont {Thorbeck}, \citenamefont {Ribeill},
  \citenamefont {Sank}, \citenamefont {Kelly}, \citenamefont {Barends},
  \citenamefont {Martinis},\ and\ \citenamefont {McDermott}}]{Hover2014}%
  \BibitemOpen
  \bibfield  {author} {\bibinfo {author} {\bibfnamefont {D.}~\bibnamefont
  {Hover}}, \bibinfo {author} {\bibfnamefont {S.}~\bibnamefont {Zhu}}, \bibinfo
  {author} {\bibfnamefont {T.}~\bibnamefont {Thorbeck}}, \bibinfo {author}
  {\bibfnamefont {G.}~\bibnamefont {Ribeill}}, \bibinfo {author} {\bibfnamefont
  {D.}~\bibnamefont {Sank}}, \bibinfo {author} {\bibfnamefont {J.}~\bibnamefont
  {Kelly}}, \bibinfo {author} {\bibfnamefont {R.}~\bibnamefont {Barends}},
  \bibinfo {author} {\bibfnamefont {J.~M.}\ \bibnamefont {Martinis}}, \ and\
  \bibinfo {author} {\bibfnamefont {R.}~\bibnamefont {McDermott}},\ }\href@noop
  {} {\bibfield  {journal} {\bibinfo  {journal} {Appl. Phys. Lett.}\ }\textbf
  {\bibinfo {volume} {104}},\ \bibinfo {pages} {152601} (\bibinfo {year}
  {2014})}\BibitemShut {NoStop}%
\bibitem [{\citenamefont {M{\"u}ck}, \citenamefont {Schmidt},\ and\
  \citenamefont {Clarke}(2017)}]{Muck2017}%
  \BibitemOpen
  \bibfield  {author} {\bibinfo {author} {\bibfnamefont {M.}~\bibnamefont
  {M{\"u}ck}}, \bibinfo {author} {\bibfnamefont {B.}~\bibnamefont {Schmidt}}, \
  and\ \bibinfo {author} {\bibfnamefont {J.}~\bibnamefont {Clarke}},\
  }\href@noop {} {\bibfield  {journal} {\bibinfo  {journal} {Appl. Phys.
  Lett.}\ }\textbf {\bibinfo {volume} {111}},\ \bibinfo {pages} {042604}
  (\bibinfo {year} {2017})}\BibitemShut {NoStop}%
\bibitem [{\citenamefont {Jones}\ \emph {et~al.}(2013)\citenamefont {Jones},
  \citenamefont {Huhtam\"{a}ki}, \citenamefont {Salmilehto}, \citenamefont
  {Tan},\ and\ \citenamefont {M\"{o}tt\"{o}nen}}]{Jones2013}%
  \BibitemOpen
  \bibfield  {author} {\bibinfo {author} {\bibfnamefont {P.~J.}\ \bibnamefont
  {Jones}}, \bibinfo {author} {\bibfnamefont {J.~A.~M.}\ \bibnamefont
  {Huhtam\"{a}ki}}, \bibinfo {author} {\bibfnamefont {J.}~\bibnamefont
  {Salmilehto}}, \bibinfo {author} {\bibfnamefont {K.~Y.}\ \bibnamefont {Tan}},
  \ and\ \bibinfo {author} {\bibfnamefont {M.}~\bibnamefont
  {M\"{o}tt\"{o}nen}},\ }\href {\doibase 10.1038/srep01987} {\bibfield
  {journal} {\bibinfo  {journal} {Sci. Rep.}\ }\textbf {\bibinfo {volume}
  {3}},\ \bibinfo {pages} {1987} (\bibinfo {year} {2013})}\BibitemShut
  {NoStop}%
\bibitem [{\citenamefont {Tuorila}\ \emph {et~al.}(2017)\citenamefont
  {Tuorila}, \citenamefont {Partanen}, \citenamefont {{Ala-Nissila}},\ and\
  \citenamefont {M\"{o}tt\"{o}nen}}]{Tuorila2017}%
  \BibitemOpen
  \bibfield  {author} {\bibinfo {author} {\bibfnamefont {J.}~\bibnamefont
  {Tuorila}}, \bibinfo {author} {\bibfnamefont {M.}~\bibnamefont {Partanen}},
  \bibinfo {author} {\bibfnamefont {T.}~\bibnamefont {{Ala-Nissila}}}, \ and\
  \bibinfo {author} {\bibfnamefont {M.}~\bibnamefont {M\"{o}tt\"{o}nen}},\
  }\href {\doibase 10.1038/s41534-017-0027-1} {\bibfield  {journal} {\bibinfo
  {journal} {npj Quantum Inf.}\ }\textbf {\bibinfo {volume} {3}},\ \bibinfo
  {pages} {27} (\bibinfo {year} {2017})}\BibitemShut {NoStop}%
\bibitem [{\citenamefont {Partanen}\ \emph {et~al.}(2018)\citenamefont
  {Partanen}, \citenamefont {Tan}, \citenamefont {Masuda}, \citenamefont
  {Govenius}, \citenamefont {Lake}, \citenamefont {Jenei}, \citenamefont
  {Gr\"{o}nberg}, \citenamefont {Hassel}, \citenamefont {Simbierowicz},
  \citenamefont {Vesterinen}, \citenamefont {Tuorila}, \citenamefont
  {{Ala-Nissila}},\ and\ \citenamefont {M{\"o}tt{\"o}nen}}]{Partanen2018}%
  \BibitemOpen
  \bibfield  {author} {\bibinfo {author} {\bibfnamefont {M.}~\bibnamefont
  {Partanen}}, \bibinfo {author} {\bibfnamefont {K.~Y.}\ \bibnamefont {Tan}},
  \bibinfo {author} {\bibfnamefont {S.}~\bibnamefont {Masuda}}, \bibinfo
  {author} {\bibfnamefont {J.}~\bibnamefont {Govenius}}, \bibinfo {author}
  {\bibfnamefont {R.~E.}\ \bibnamefont {Lake}}, \bibinfo {author}
  {\bibfnamefont {M.}~\bibnamefont {Jenei}}, \bibinfo {author} {\bibfnamefont
  {L.}~\bibnamefont {Gr\"{o}nberg}}, \bibinfo {author} {\bibfnamefont
  {J.}~\bibnamefont {Hassel}}, \bibinfo {author} {\bibfnamefont
  {S.}~\bibnamefont {Simbierowicz}}, \bibinfo {author} {\bibfnamefont
  {V.}~\bibnamefont {Vesterinen}}, \bibinfo {author} {\bibfnamefont
  {J.}~\bibnamefont {Tuorila}}, \bibinfo {author} {\bibfnamefont
  {T.}~\bibnamefont {{Ala-Nissila}}}, \ and\ \bibinfo {author} {\bibfnamefont
  {M.}~\bibnamefont {M{\"o}tt{\"o}nen}},\ }\href {\doibase
  10.1038/s41598-018-24449-1} {\bibfield  {journal} {\bibinfo  {journal} {Sci.
  Rep.}\ }\textbf {\bibinfo {volume} {8}},\ \bibinfo {pages} {6325} (\bibinfo
  {year} {2018})}\BibitemShut {NoStop}%
\bibitem [{\citenamefont {Verstraete}, \citenamefont {Wolf},\ and\
  \citenamefont {Cirac}(2009)}]{Verstraete2009}%
  \BibitemOpen
  \bibfield  {author} {\bibinfo {author} {\bibfnamefont {F.}~\bibnamefont
  {Verstraete}}, \bibinfo {author} {\bibfnamefont {M.~M.}\ \bibnamefont
  {Wolf}}, \ and\ \bibinfo {author} {\bibfnamefont {J.~I.}\ \bibnamefont
  {Cirac}},\ }\href@noop {} {\bibfield  {journal} {\bibinfo  {journal} {Nature
  physics}\ }\textbf {\bibinfo {volume} {5}},\ \bibinfo {pages} {633} (\bibinfo
  {year} {2009})}\BibitemShut {NoStop}%
\bibitem [{\citenamefont {Niskanen}, \citenamefont {Nakamura},\ and\
  \citenamefont {Pekola}(2007)}]{Niskanen2007}%
  \BibitemOpen
  \bibfield  {author} {\bibinfo {author} {\bibfnamefont {A.~O.}\ \bibnamefont
  {Niskanen}}, \bibinfo {author} {\bibfnamefont {Y.}~\bibnamefont {Nakamura}},
  \ and\ \bibinfo {author} {\bibfnamefont {J.~P.}\ \bibnamefont {Pekola}},\
  }\href {\doibase 10.1103/PhysRevB.76.174523} {\bibfield  {journal} {\bibinfo
  {journal} {Phys. Rev. B}\ }\textbf {\bibinfo {volume} {76}},\ \bibinfo
  {pages} {174523} (\bibinfo {year} {2007})}\BibitemShut {NoStop}%
\bibitem [{\citenamefont {Karimi}\ and\ \citenamefont
  {Pekola}(2016)}]{Karimi2016}%
  \BibitemOpen
  \bibfield  {author} {\bibinfo {author} {\bibfnamefont {B.}~\bibnamefont
  {Karimi}}\ and\ \bibinfo {author} {\bibfnamefont {J.~P.}\ \bibnamefont
  {Pekola}},\ }\href {\doibase 10.1103/PhysRevB.94.184503} {\bibfield
  {journal} {\bibinfo  {journal} {Phys. Rev. B}\ }\textbf {\bibinfo {volume}
  {94}},\ \bibinfo {pages} {184503} (\bibinfo {year} {2016})}\BibitemShut
  {NoStop}%
\bibitem [{\citenamefont {Ronzani}\ \emph {et~al.}(2018)\citenamefont
  {Ronzani}, \citenamefont {Karimi}, \citenamefont {Senior}, \citenamefont
  {Chang}, \citenamefont {Peltonen}, \citenamefont {Chen},\ and\ \citenamefont
  {Pekola}}]{Ronzani2018}%
  \BibitemOpen
  \bibfield  {author} {\bibinfo {author} {\bibfnamefont {A.}~\bibnamefont
  {Ronzani}}, \bibinfo {author} {\bibfnamefont {B.}~\bibnamefont {Karimi}},
  \bibinfo {author} {\bibfnamefont {J.}~\bibnamefont {Senior}}, \bibinfo
  {author} {\bibfnamefont {Y.-C.}\ \bibnamefont {Chang}}, \bibinfo {author}
  {\bibfnamefont {J.~T.}\ \bibnamefont {Peltonen}}, \bibinfo {author}
  {\bibfnamefont {C.}~\bibnamefont {Chen}}, \ and\ \bibinfo {author}
  {\bibfnamefont {J.~P.}\ \bibnamefont {Pekola}},\ }\href@noop {} {\bibfield
  {journal} {\bibinfo  {journal} {Nature Physics}\ }\textbf {\bibinfo {volume}
  {14}},\ \bibinfo {pages} {991} (\bibinfo {year} {2018})}\BibitemShut
  {NoStop}%
\bibitem [{\citenamefont {Nahum}, \citenamefont {Eiles},\ and\ \citenamefont
  {Martinis}(1994)}]{Nahum1994}%
  \BibitemOpen
  \bibfield  {author} {\bibinfo {author} {\bibfnamefont {M.}~\bibnamefont
  {Nahum}}, \bibinfo {author} {\bibfnamefont {T.~M.}\ \bibnamefont {Eiles}}, \
  and\ \bibinfo {author} {\bibfnamefont {J.~M.}\ \bibnamefont {Martinis}},\
  }\href {\doibase 10.1063/1.112456} {\bibfield  {journal} {\bibinfo  {journal}
  {Appl. Phys. Lett.}\ }\textbf {\bibinfo {volume} {65}},\ \bibinfo {pages}
  {3123} (\bibinfo {year} {1994})}\BibitemShut {NoStop}%
\bibitem [{\citenamefont {Giazotto}\ \emph {et~al.}(2006)\citenamefont
  {Giazotto}, \citenamefont {Heikkil\"{a}}, \citenamefont {Luukanen},
  \citenamefont {Savin},\ and\ \citenamefont {Pekola}}]{Giazotto2006}%
  \BibitemOpen
  \bibfield  {author} {\bibinfo {author} {\bibfnamefont {F.}~\bibnamefont
  {Giazotto}}, \bibinfo {author} {\bibfnamefont {T.~T.}\ \bibnamefont
  {Heikkil\"{a}}}, \bibinfo {author} {\bibfnamefont {A.}~\bibnamefont
  {Luukanen}}, \bibinfo {author} {\bibfnamefont {A.~M.}\ \bibnamefont {Savin}},
  \ and\ \bibinfo {author} {\bibfnamefont {J.~P.}\ \bibnamefont {Pekola}},\
  }\href {\doibase 10.1103/RevModPhys.78.217} {\bibfield  {journal} {\bibinfo
  {journal} {Rev. Mod. Phys.}\ }\textbf {\bibinfo {volume} {78}},\ \bibinfo
  {pages} {217} (\bibinfo {year} {2006})}\BibitemShut {NoStop}%
\bibitem [{\citenamefont {Gasparinetti}\ \emph {et~al.}(2015)\citenamefont
  {Gasparinetti}, \citenamefont {Viisanen}, \citenamefont {Saira},
  \citenamefont {Faivre}, \citenamefont {Arzeo}, \citenamefont {Meschke},\ and\
  \citenamefont {Pekola}}]{Gasparinetti2015}%
  \BibitemOpen
  \bibfield  {author} {\bibinfo {author} {\bibfnamefont {S.}~\bibnamefont
  {Gasparinetti}}, \bibinfo {author} {\bibfnamefont {K.~L.}\ \bibnamefont
  {Viisanen}}, \bibinfo {author} {\bibfnamefont {O.-P.}\ \bibnamefont {Saira}},
  \bibinfo {author} {\bibfnamefont {T.}~\bibnamefont {Faivre}}, \bibinfo
  {author} {\bibfnamefont {M.}~\bibnamefont {Arzeo}}, \bibinfo {author}
  {\bibfnamefont {M.}~\bibnamefont {Meschke}}, \ and\ \bibinfo {author}
  {\bibfnamefont {J.~P.}\ \bibnamefont {Pekola}},\ }\href {\doibase
  10.1103/PhysRevApplied.3.014007} {\bibfield  {journal} {\bibinfo  {journal}
  {Phys. Rev. Appl.}\ }\textbf {\bibinfo {volume} {3}},\ \bibinfo {pages}
  {014007} (\bibinfo {year} {2015})}\BibitemShut {NoStop}%
\bibitem [{\citenamefont {Saira}\ \emph {et~al.}(2016)\citenamefont {Saira},
  \citenamefont {Zgirski}, \citenamefont {Viisanen}, \citenamefont {Golubev},\
  and\ \citenamefont {Pekola}}]{Saira2016}%
  \BibitemOpen
  \bibfield  {author} {\bibinfo {author} {\bibfnamefont {O.-P.}\ \bibnamefont
  {Saira}}, \bibinfo {author} {\bibfnamefont {M.}~\bibnamefont {Zgirski}},
  \bibinfo {author} {\bibfnamefont {K.~L.}\ \bibnamefont {Viisanen}}, \bibinfo
  {author} {\bibfnamefont {D.~S.}\ \bibnamefont {Golubev}}, \ and\ \bibinfo
  {author} {\bibfnamefont {J.~P.}\ \bibnamefont {Pekola}},\ }\href {\doibase
  10.1103/PhysRevApplied.6.024005} {\bibfield  {journal} {\bibinfo  {journal}
  {Phys. Rev. Appl.}\ }\textbf {\bibinfo {volume} {6}},\ \bibinfo {pages}
  {024005} (\bibinfo {year} {2016})}\BibitemShut {NoStop}%
\bibitem [{\citenamefont {Masuda}\ \emph {et~al.}(2018)\citenamefont {Masuda},
  \citenamefont {Tan}, \citenamefont {Partanen}, \citenamefont {Lake},
  \citenamefont {Govenius}, \citenamefont {Silveri}, \citenamefont {Grabert},\
  and\ \citenamefont {M\"{o}tt\"{o}nen}}]{Masuda2018}%
  \BibitemOpen
  \bibfield  {author} {\bibinfo {author} {\bibfnamefont {S.}~\bibnamefont
  {Masuda}}, \bibinfo {author} {\bibfnamefont {K.~Y.}\ \bibnamefont {Tan}},
  \bibinfo {author} {\bibfnamefont {M.}~\bibnamefont {Partanen}}, \bibinfo
  {author} {\bibfnamefont {R.~E.}\ \bibnamefont {Lake}}, \bibinfo {author}
  {\bibfnamefont {J.}~\bibnamefont {Govenius}}, \bibinfo {author}
  {\bibfnamefont {M.}~\bibnamefont {Silveri}}, \bibinfo {author} {\bibfnamefont
  {H.}~\bibnamefont {Grabert}}, \ and\ \bibinfo {author} {\bibfnamefont
  {M.}~\bibnamefont {M\"{o}tt\"{o}nen}},\ }\href {\doibase
  10.1038/s41598-018-21772-5} {\bibfield  {journal} {\bibinfo  {journal} {Sci.
  Rep.}\ }\textbf {\bibinfo {volume} {8}},\ \bibinfo {pages} {3966} (\bibinfo
  {year} {2018})}\BibitemShut {NoStop}%
\bibitem [{\citenamefont {George}\ \emph {et~al.}(2017)\citenamefont {George},
  \citenamefont {Senior}, \citenamefont {Saira}, \citenamefont {Pekola},
  \citenamefont {{de Graaf}}, \citenamefont {Lindstr\"{o}m},\ and\
  \citenamefont {Pashkin}}]{George2017}%
  \BibitemOpen
  \bibfield  {author} {\bibinfo {author} {\bibfnamefont {R.~E.}\ \bibnamefont
  {George}}, \bibinfo {author} {\bibfnamefont {J.}~\bibnamefont {Senior}},
  \bibinfo {author} {\bibfnamefont {O.-P.}\ \bibnamefont {Saira}}, \bibinfo
  {author} {\bibfnamefont {J.~P.}\ \bibnamefont {Pekola}}, \bibinfo {author}
  {\bibfnamefont {S.~E.}\ \bibnamefont {{de Graaf}}}, \bibinfo {author}
  {\bibfnamefont {T.}~\bibnamefont {Lindstr\"{o}m}}, \ and\ \bibinfo {author}
  {\bibfnamefont {Y.~A.}\ \bibnamefont {Pashkin}},\ }\href {\doibase
  10.1007/s10909-017-1787-x} {\bibfield  {journal} {\bibinfo  {journal} {J. Low
  Temp. Phys.}\ }\textbf {\bibinfo {volume} {189}},\ \bibinfo {pages} {60}
  (\bibinfo {year} {2017})}\BibitemShut {NoStop}%
\bibitem [{\citenamefont {Urick}, \citenamefont {Williams},\ and\ \citenamefont
  {McKinney}(2015)}]{Urick2015}%
  \BibitemOpen
  \bibfield  {author} {\bibinfo {author} {\bibfnamefont {V.~J.}\ \bibnamefont
  {Urick}}, \bibinfo {author} {\bibfnamefont {K.~J.}\ \bibnamefont {Williams}},
  \ and\ \bibinfo {author} {\bibfnamefont {J.~D.}\ \bibnamefont {McKinney}},\
  }\href@noop {} {\emph {\bibinfo {title} {Fundamentals of Microwave
  Photonics}}},\ Vol.~\bibinfo {volume} {1}\ (\bibinfo  {publisher} {{John
  Wiley \& Sons}},\ \bibinfo {year} {2015})\BibitemShut {NoStop}%
\bibitem [{\citenamefont {Khalil}\ \emph {et~al.}(2012)\citenamefont {Khalil},
  \citenamefont {Stoutimore}, \citenamefont {Wellstood},\ and\ \citenamefont
  {Osborn}}]{Khalil2012}%
  \BibitemOpen
  \bibfield  {author} {\bibinfo {author} {\bibfnamefont {M.~S.}\ \bibnamefont
  {Khalil}}, \bibinfo {author} {\bibfnamefont {M.~J.~A.}\ \bibnamefont
  {Stoutimore}}, \bibinfo {author} {\bibfnamefont {F.~C.}\ \bibnamefont
  {Wellstood}}, \ and\ \bibinfo {author} {\bibfnamefont {K.~D.}\ \bibnamefont
  {Osborn}},\ }\href {\doibase 10.1063/1.3692073} {\bibfield  {journal}
  {\bibinfo  {journal} {J. Appl. Phys.}\ }\textbf {\bibinfo {volume} {111}},\
  \bibinfo {pages} {054510} (\bibinfo {year} {2012})}\BibitemShut {NoStop}%
\bibitem [{\citenamefont {Deng}, \citenamefont {Otto},\ and\ \citenamefont
  {Lupascu}(2013)}]{Deng2013}%
  \BibitemOpen
  \bibfield  {author} {\bibinfo {author} {\bibfnamefont {C.}~\bibnamefont
  {Deng}}, \bibinfo {author} {\bibfnamefont {M.}~\bibnamefont {Otto}}, \ and\
  \bibinfo {author} {\bibfnamefont {A.}~\bibnamefont {Lupascu}},\ }\href
  {\doibase 10.1063/1.4817512} {\bibfield  {journal} {\bibinfo  {journal} {J.
  Appl. Phys.}\ }\textbf {\bibinfo {volume} {114}},\ \bibinfo {pages} {054504}
  (\bibinfo {year} {2013})}\BibitemShut {NoStop}%
\bibitem [{\citenamefont {Probst}\ \emph {et~al.}(2015)\citenamefont {Probst},
  \citenamefont {Song}, \citenamefont {Bushev}, \citenamefont {Ustinov},\ and\
  \citenamefont {Weides}}]{Probst2015}%
  \BibitemOpen
  \bibfield  {author} {\bibinfo {author} {\bibfnamefont {S.}~\bibnamefont
  {Probst}}, \bibinfo {author} {\bibfnamefont {F.~B.}\ \bibnamefont {Song}},
  \bibinfo {author} {\bibfnamefont {P.~A.}\ \bibnamefont {Bushev}}, \bibinfo
  {author} {\bibfnamefont {A.~V.}\ \bibnamefont {Ustinov}}, \ and\ \bibinfo
  {author} {\bibfnamefont {M.}~\bibnamefont {Weides}},\ }\href {\doibase
  10.1063/1.4907935} {\bibfield  {journal} {\bibinfo  {journal} {Rev. Sci.
  Instrum}\ }\textbf {\bibinfo {volume} {86}},\ \bibinfo {pages} {024706}
  (\bibinfo {year} {2015})}\BibitemShut {NoStop}%
\bibitem [{\citenamefont {G\"{o}ppl}\ \emph {et~al.}(2008)\citenamefont
  {G\"{o}ppl}, \citenamefont {Fragner}, \citenamefont {Baur}, \citenamefont
  {Bianchetti}, \citenamefont {Filipp}, \citenamefont {Fink}, \citenamefont
  {Leek}, \citenamefont {Puebla}, \citenamefont {Steffen},\ and\ \citenamefont
  {Wallraff}}]{Goppl2008}%
  \BibitemOpen
  \bibfield  {author} {\bibinfo {author} {\bibfnamefont {M.}~\bibnamefont
  {G\"{o}ppl}}, \bibinfo {author} {\bibfnamefont {A.}~\bibnamefont {Fragner}},
  \bibinfo {author} {\bibfnamefont {M.}~\bibnamefont {Baur}}, \bibinfo {author}
  {\bibfnamefont {R.}~\bibnamefont {Bianchetti}}, \bibinfo {author}
  {\bibfnamefont {S.}~\bibnamefont {Filipp}}, \bibinfo {author} {\bibfnamefont
  {J.~M.}\ \bibnamefont {Fink}}, \bibinfo {author} {\bibfnamefont {P.~J.}\
  \bibnamefont {Leek}}, \bibinfo {author} {\bibfnamefont {G.}~\bibnamefont
  {Puebla}}, \bibinfo {author} {\bibfnamefont {L.}~\bibnamefont {Steffen}}, \
  and\ \bibinfo {author} {\bibfnamefont {A.}~\bibnamefont {Wallraff}},\ }\href
  {\doibase 10.1063/1.3010859} {\bibfield  {journal} {\bibinfo  {journal} {J.
  Appl. Phys.}\ }\textbf {\bibinfo {volume} {104}},\ \bibinfo {pages} {113904}
  (\bibinfo {year} {2008})}\BibitemShut {NoStop}%
\bibitem [{\citenamefont {Meservey}\ and\ \citenamefont
  {Tedrow}(1969)}]{Meservey1969}%
  \BibitemOpen
  \bibfield  {author} {\bibinfo {author} {\bibfnamefont {R.}~\bibnamefont
  {Meservey}}\ and\ \bibinfo {author} {\bibfnamefont {P.~M.}\ \bibnamefont
  {Tedrow}},\ }\href {\doibase 10.1063/1.1657905} {\bibfield  {journal}
  {\bibinfo  {journal} {J. Appl. Phys.}\ }\textbf {\bibinfo {volume} {40}},\
  \bibinfo {pages} {2028} (\bibinfo {year} {1969})}\BibitemShut {NoStop}%
\bibitem [{\citenamefont {Inomata}\ \emph {et~al.}(2009)\citenamefont
  {Inomata}, \citenamefont {Yamamoto}, \citenamefont {Watanabe}, \citenamefont
  {Matsuba},\ and\ \citenamefont {Tsai}}]{Inomata2009}%
  \BibitemOpen
  \bibfield  {author} {\bibinfo {author} {\bibfnamefont {K.}~\bibnamefont
  {Inomata}}, \bibinfo {author} {\bibfnamefont {T.}~\bibnamefont {Yamamoto}},
  \bibinfo {author} {\bibfnamefont {M.}~\bibnamefont {Watanabe}}, \bibinfo
  {author} {\bibfnamefont {K.}~\bibnamefont {Matsuba}}, \ and\ \bibinfo
  {author} {\bibfnamefont {J.-S.}\ \bibnamefont {Tsai}},\ }\href {\doibase
  10.1116/1.3232301} {\bibfield  {journal} {\bibinfo  {journal} {J. Vac. Sci.
  Technol.}\ }\textbf {\bibinfo {volume} {27}},\ \bibinfo {pages} {2286}
  (\bibinfo {year} {2009})}\BibitemShut {NoStop}%
\bibitem [{\citenamefont {Wei}\ \emph {et~al.}(2011)\citenamefont {Wei},
  \citenamefont {{Cadden-Zimansky}}, \citenamefont {Chandrasekhar},\ and\
  \citenamefont {Virtanen}}]{Wei2011}%
  \BibitemOpen
  \bibfield  {author} {\bibinfo {author} {\bibfnamefont {J.}~\bibnamefont
  {Wei}}, \bibinfo {author} {\bibfnamefont {P.}~\bibnamefont
  {{Cadden-Zimansky}}}, \bibinfo {author} {\bibfnamefont {V.}~\bibnamefont
  {Chandrasekhar}}, \ and\ \bibinfo {author} {\bibfnamefont {P.}~\bibnamefont
  {Virtanen}},\ }\href {\doibase 10.1103/PhysRevB.84.224519} {\bibfield
  {journal} {\bibinfo  {journal} {Phys. Rev. B}\ }\textbf {\bibinfo {volume}
  {84}},\ \bibinfo {pages} {224519} (\bibinfo {year} {2011})}\BibitemShut
  {NoStop}%
\bibitem [{\citenamefont {Angers}\ \emph {et~al.}(2008)\citenamefont {Angers},
  \citenamefont {Chiodi}, \citenamefont {Montambaux}, \citenamefont {Ferrier},
  \citenamefont {Gu\'{e}ron}, \citenamefont {Bouchiat},\ and\ \citenamefont
  {Cuevas}}]{Angers2008}%
  \BibitemOpen
  \bibfield  {author} {\bibinfo {author} {\bibfnamefont {L.}~\bibnamefont
  {Angers}}, \bibinfo {author} {\bibfnamefont {F.}~\bibnamefont {Chiodi}},
  \bibinfo {author} {\bibfnamefont {G.}~\bibnamefont {Montambaux}}, \bibinfo
  {author} {\bibfnamefont {M.}~\bibnamefont {Ferrier}}, \bibinfo {author}
  {\bibfnamefont {S.}~\bibnamefont {Gu\'{e}ron}}, \bibinfo {author}
  {\bibfnamefont {H.}~\bibnamefont {Bouchiat}}, \ and\ \bibinfo {author}
  {\bibfnamefont {J.~C.}\ \bibnamefont {Cuevas}},\ }\href {\doibase
  10.1103/PhysRevB.77.165408} {\bibfield  {journal} {\bibinfo  {journal} {Phys.
  Rev. B}\ }\textbf {\bibinfo {volume} {77}},\ \bibinfo {pages} {165408}
  (\bibinfo {year} {2008})}\BibitemShut {NoStop}%
\bibitem [{\citenamefont {Clarke}(1969)}]{Clarke1969}%
  \BibitemOpen
  \bibfield  {author} {\bibinfo {author} {\bibfnamefont {J.}~\bibnamefont
  {Clarke}},\ }\href {\doibase 10.1098/rspa.1969.0020} {\bibfield  {journal}
  {\bibinfo  {journal} {Proc. R. Soc. Lond. A}\ }\textbf {\bibinfo {volume}
  {308}},\ \bibinfo {pages} {447} (\bibinfo {year} {1969})}\BibitemShut
  {NoStop}%
\bibitem [{\citenamefont {Shepherd}(1972)}]{Shepherd1972}%
  \BibitemOpen
  \bibfield  {author} {\bibinfo {author} {\bibfnamefont {J.~G.}\ \bibnamefont
  {Shepherd}},\ }\href {\doibase 10.1098/rspa.1972.0018} {\bibfield  {journal}
  {\bibinfo  {journal} {Proc. R. Soc. Lond. A}\ }\textbf {\bibinfo {volume}
  {326}},\ \bibinfo {pages} {421} (\bibinfo {year} {1972})}\BibitemShut
  {NoStop}%
\bibitem [{\citenamefont {{de Gennes}}(1964)}]{DeGennes1964}%
  \BibitemOpen
  \bibfield  {author} {\bibinfo {author} {\bibfnamefont {P.~G.}\ \bibnamefont
  {{de Gennes}}},\ }\href {\doibase 10.1103/RevModPhys.36.225} {\bibfield
  {journal} {\bibinfo  {journal} {Rev. Mod. Phys.}\ }\textbf {\bibinfo {volume}
  {36}},\ \bibinfo {pages} {225} (\bibinfo {year} {1964})}\BibitemShut
  {NoStop}%
\bibitem [{\citenamefont {Dubos}\ \emph {et~al.}(2001)\citenamefont {Dubos},
  \citenamefont {Courtois}, \citenamefont {Pannetier}, \citenamefont {Wilhelm},
  \citenamefont {Zaikin},\ and\ \citenamefont {Sch\"{o}n}}]{Dubos2001}%
  \BibitemOpen
  \bibfield  {author} {\bibinfo {author} {\bibfnamefont {P.}~\bibnamefont
  {Dubos}}, \bibinfo {author} {\bibfnamefont {H.}~\bibnamefont {Courtois}},
  \bibinfo {author} {\bibfnamefont {B.}~\bibnamefont {Pannetier}}, \bibinfo
  {author} {\bibfnamefont {F.~K.}\ \bibnamefont {Wilhelm}}, \bibinfo {author}
  {\bibfnamefont {A.~D.}\ \bibnamefont {Zaikin}}, \ and\ \bibinfo {author}
  {\bibfnamefont {G.}~\bibnamefont {Sch\"{o}n}},\ }\href {\doibase
  10.1103/PhysRevB.63.064502} {\bibfield  {journal} {\bibinfo  {journal} {Phys.
  Rev. B}\ }\textbf {\bibinfo {volume} {63}},\ \bibinfo {pages} {064502}
  (\bibinfo {year} {2001})}\BibitemShut {NoStop}%
\bibitem [{\citenamefont {Andreev}(1965)}]{Andreev1965}%
  \BibitemOpen
  \bibfield  {author} {\bibinfo {author} {\bibfnamefont {A.}~\bibnamefont
  {Andreev}},\ }\href@noop {} {\bibfield  {journal} {\bibinfo  {journal} {Sov.
  Phys. JETP}\ }\textbf {\bibinfo {volume} {20}},\ \bibinfo {pages} {1490}
  (\bibinfo {year} {1965})}\BibitemShut {NoStop}%
\bibitem [{\citenamefont {Virtanen}\ \emph {et~al.}(2011)\citenamefont
  {Virtanen}, \citenamefont {Bergeret}, \citenamefont {Cuevas},\ and\
  \citenamefont {Heikkil\"{a}}}]{Virtanen2011}%
  \BibitemOpen
  \bibfield  {author} {\bibinfo {author} {\bibfnamefont {P.}~\bibnamefont
  {Virtanen}}, \bibinfo {author} {\bibfnamefont {F.~S.}\ \bibnamefont
  {Bergeret}}, \bibinfo {author} {\bibfnamefont {J.~C.}\ \bibnamefont
  {Cuevas}}, \ and\ \bibinfo {author} {\bibfnamefont {T.~T.}\ \bibnamefont
  {Heikkil\"{a}}},\ }\href {\doibase 10.1103/PhysRevB.83.144514} {\bibfield
  {journal} {\bibinfo  {journal} {Phys. Rev. B}\ }\textbf {\bibinfo {volume}
  {83}},\ \bibinfo {pages} {144514} (\bibinfo {year} {2011})}\BibitemShut
  {NoStop}%
\end{thebibliography}%

\end{document}